# Terrestrial effects of possible astrophysical sources of an AD 774-775 increase in $^{14}$C production

Brian C. Thomas[1], Adrian L. Melott[2], Keith R. Arkenberg[1], and Brock R. Snyder II[1]

*1 Department of Physics and Astronomy, Washburn University, Topeka, Kansas 66621 USA* E-mail: brian.thomas@washburn.edu

*2 Department of Physics and Astronomy, University of Kansas, Lawrence, Kansas 66045 USA.*

**ABSTRACT**

**We examine possible sources of a substantial increase in tree ring $^{14}$C measurements for the years AD 774-775. Contrary to claims regarding a coronal mass ejection (CME), the required CME energy is not several orders of magnitude greater than known solar events. We consider solar proton events (SPEs) with three different fluences and two different spectra. The data may be explained by an event with fluence about one order of magnitude beyond the October 1989 SPE. Two hard spectrum cases considered here result in moderate ozone depletion, so no mass extinction is implied, though we do predict increases in erythema and damage to plants from enhanced solar UV. We are able to rule out an event with a very soft spectrum that causes severe ozone depletion and subsequent biological impacts. Nitrate enhancements are consistent with their apparent absence in ice core data. The modern technological implications of such an event may be extreme, and considering recent confirmation of superflares on solar-type stars, this issue merits attention.**

## 1. Introduction

*Miyake et al* [2012] found that tree ring data imply a strong increase in the rate of production of $^{14}C$ during AD 774-775. They put it in the context of supernovae and solar proton events. We extend that consideration here. Additional candidate extrasolar events lie within the class of unidentified or poorly-understood high-energy transients [*Gehrels and Cannizzo,* 2012], some of which made measurable modifications to the ionization state of the Earth's atmosphere. A short gamma-ray burst (hereafter GRB) may have provided the source, but the probability of an event with the terrestrial fluence as estimated by *Miyake et al.* [2012] is about $10^{-7}$ y$^{-1}$ [*Melott and Thomas,* 2011], so the probability of an event in the last 1250 years is about $10^{-4}$ [*Melott and Thomas,* 2012]. *Hambaryan and Neuhaeuser* [2013] emphasize uncertainties in the short-burst GRB rate estimates. *Eichler and Mordecai* [2013] suggest a solar comet impact, but without questioning the Miyake energy scaling, which we have argued is unnecessary and unlikely [*Melott and Thomas,* 2012].

*Miyake et al.* [2012] concluded that a recent supernova near enough to produce this effect should have left a conspicuous remnant. We agree. They further argued that the energy implied for a solar proton event (SPE) was many orders of magnitude beyond anything known on the Sun [*Schaeffer,* 2012], and therefore implausible. We disagree.

The scaling from the SPE to the CME energy in *Miyake et al.* [2012] was incorrect [*Melott and Thomas,* 2012]. They assumed CMEs propagated isotropically, but they are primarily confined to opening angles of 24° to 72° [*Bothmer and Zhukov,* 2007], with smaller angles much more common [*Schrijver,* 2010]. The few wide angle CMEs are much less likely (per steradian) to produce an SPE [*Park et al.*, 2012], suggesting considerably lower fluences. Such "global" CMEs are rare, and account for many fewer SPEs than implied by their angular coverage. The largest observed SPEs have not corresponded to wide angle CMEs. Assuming 0.1 steradian, the implied CME energy is reduced below the *Miyake et al.* [2012] estimate by two orders of magnitude to about 2 x $10^{26}$ J. The energy deposition at the Earth estimated by *Miyake et al.* [2012] in 30 MeV units and divided by the surface area is 3 x $10^{11}$ protons cm$^{-2}$, about 71 times the fluence of the October 1989 SPE [*McCracken et al.*, 2001], one of the most intense SPEs in modern times.

*Usoskin and Kovaltsov* [2012] estimated the fluence of the 774 AD event to be about 3 x $10^{10}$ protons cm$^{-2}$, assuming a spectrum associated with the extremely hard 1956 SPE. This is about 30 times their estimate of the time-averaged annual solar energetic particle fluence. Their substantial differences with *Miyake et al.* [2012] on the estimated fluence are based on the choice of fitting spectrum—one is soft, the other very hard. They state that a soft spectrum like that of the August 1972 SPE may have a fluence 40 times greater than their 1956 (extremely hard) spectrum taken as a standard in order to produce the same $^{14}$C increase. As there is no independent data on pre-space era events, we will deal with this uncertainty by conducting simulations of atmospheric affects using the October 1989 SPE spectrum with two different E > 30 MeV proton fluence values, as well as a third fluence case using the August 1972 SPE spectrum. The October 1989 spectrum is relatively hard, but lies between the very soft 1972 SPE and the very hard 1956 SPE.

$^{10}$Be enhancements may also be expected. However, they are strongly precipitation dependent, possibly connected more to local weather than to atmospheric concentrations, and reliably correlated with $^{14}$C only for centennial scale variation [*Finkel and Nishiizumi* 1997]. Transport to Greenland and the Antarctic are fundamentally different, and peak at different times of year [*Pedro et al.* 2011a], possibly only reflecting cosmic ray activity in the previous 10 months [*Pedro et al.* 2011b]. They also show extreme geographic variability [*Berggren et al.* 2012] and are therefore suspect as compared with $^{14}$C. *Beer et al.* [2012] show that they are useful and reliable for enhancements which arise on a timescale of 10 years or longer, which is not useful for SPEs or possible enhancement from a comet impact [*Overholt and Melott*, manuscript in preparation, 2012].

## 2. SPE and stellar flare event rates

The probability of such an event can be studied using the statistics of rare events [*Love,* 2012]. Using the 774 AD event as an example, we must ask when such events occurred earlier. *Usoskin and Kovaltsov* [2012] argue based on the $^{14}$C record that there were no such events for 11,000 years. We are primarily interested in the fluence for purposes of atmospheric ionization, and as noted by them, a softer (1972) type spectrum with the same fluence would deposit 40 times less $^{14}$C. Therefore we regard constraints on earlier possible events as not very reliable. The best we can do is use the frequency of events that show up in the $^{14}$C record, and then compare it with best-estimates from a variety of terrestrial radionuclides [*Schrijver* et al. 2012] and stellar [*Maehara* et al. 2012] records. *Love* [2012] uses both Bayesian and conventional frequentist statistics to derive rate probabilities and their uncertainties from rare or even single events. The two methods give nearly identical results, and the primary effect of small numbers of events is, as expected, larger uncertainty in the rate estimates. Assuming only the information that there was just one such event within ~1250 years, 1σ confidence intervals on the mean rate lie from 2 x 10$^{-4}$ yr$^{-1}$ to 1.8 x 10$^{-3}$ yr$^{-1}$. Using the best estimate of the mean rate of such an event, 8 x 10$^{-4}$ yr$^{-1}$, the probability of such an event within the next decade is 0.8%, which may be considered as small. However, this is insufficiently small for safety in view of such a threatening event. We note that probabilities based on *Usoskin and Kovaltsov* [2012] would be about an order of magnitude lower, but these would be probabilities of equivalent $^{14}$C production, not of equivalent proton fluence.

*Schrijver et al.* [2012] recently evaluated the frequency of extremely energetic solar events from a variety of records. We find that the event rates and energetics we propose here are consistent with their CME and flare energy rates and proposed extrapolations in their Figures 2 and 3.

We can also compare our results with newly available estimates of the rate of flares on solar-type stars. There is a rough correlation with a great deal of scatter between flare energy and SPE energy. *Maehara et al.* [2012] use bolometric luminosity (visible light for G-type main sequence stars), to estimate the rate of occurrence for superflares (E > 3 x $10^{26}$ J) for slowly rotating (T > 10 d) stars with surface temperatures close to that of the Sun to be 2.9 x $10^{-3}$ $yr^{-1}$. This is greater than the rate we estimate, but within 2σ of it. However, they estimate their detection efficiency to be impaired at this, the lower bound of their detectable flare energy, implying a likely rate much larger than we estimate above. They suggest a rate power-law for these higher-energy events of order $E^{-2}$, which is slightly steeper than that for lower-energy Solar flares. Their observations support no role for "hot Jupiters" in triggering the observed events, so the Sun remains a candidate for superflares similar to those observed on other G-type main sequence stars. Extrapolating their rates to high energies suggests moderate extinction events (see below) every million years or so.

### 3. A possible eighth century event in context

Solar flares are emission of photons, usually including and up through X-ray energies. They are often associated with CMEs, but it is possible to have each without the other [e.g. *Gosling* 1993]. In recent large events the total of magnetic, kinetic, and total radiated energy is on average a bit more than twice the kinetic energy [*Emslie et al.* 2004, 2005, 2012], so that flare energies are typically roughly comparable to CME proton energies. For the proposed AD 774-775 event, our corrected scaling from the *Miyake et al.* [2012] fluence estimates implies that the proposed CME energy is within the lower range of flare energies for solar-type stars [*Schaeffer et al.,* 2000; *Maehara et al.,* 2012], which ranges from $10^{26}$-$10^{33}$ J. *Usoskin and Kovaltsov* [2012] made a lower estimate using the 1956 event, an extreme outlier with a hard spectrum. Clearly there are sunlike stars with energy available to push CMEs many orders of magnitude beyond that implied for an AD 774-775 event. Given the poor constraints on the rates of such events at the Sun [*Melott and Thomas,* 2011; *Reedy,* 1996], it would be wise to consider the possibility. An event significantly more energetic than that of 1989 would be a disaster for electromagnetic technology [*NRC Space Studies Board,* 2008], causing widespread damage to satellites as well as transformers linking the power grid.

### 4. The SPE threat in view of the data

We have emphasized the hazard to our electromagnetic technological infrastructure. Here we consider effects on the biota. Atmospheric ionization by charged particles makes possible large increases in stratospheric $NO_x$ (N, NO, $NO_2$), which depletes ozone and allows increases in the dangerous solar UVB that reaches the ground [*Thomas et al.,* 2007; *Ejzak et al.*, 2007]. *Miyake et al.* [2012] argue against the superflare hypothesis by noting that there was no mass extinction at that time.

We now provide information to assess the biological consequences by computing the changes in atmospheric chemistry induced by the irradiation implied by their $^{14}C$ computations. We used the Goddard Space Flight Center two-dimensional (latitude, altitude) time-dependent atmospheric model that has been used extensively to model the effects of solar flares, as well as supernovae and GRB burst effects. Space constraints forbid a full description, but extensive accounts are given elsewhere [*Thomas et al.* 2005, 2007; *Ejzak et al.* 2007]. Atmospheric ionization is computed following methods described in *Thomas et al.* [2005], *Rodger et al.* [2008], *Verronen et al.* [2005], and *Jackman et al.* [1980]. All proton fluences discussed below are for 30 MeV < E < 500 MeV. The ionization method used here becomes inaccurate below an altitude of about 20 km. However, we are interested primarily in changes in total column density $O_3$, so this is not a major uncertainty for the results presented here.

We have modelled three SPE cases. For two cases we use the relatively hard spectrum associated with the SPE of 19 October 1989 [*Rodger et al.* 2008]. *Usoskin and Kovaltsov* [2012] infer a fluence of 3 x $10^{10}$ protons $cm^{-2}$ for the extremely hard 1956 SPE spectrum. We use this fluence as a lower bound, since the 1989 spectrum is not as hard and a higher fluence is required to get the same radiocarbon yield. As a second case we use the 1989 spectrum with a fluence of 3 x $10^{11}$ protons $cm^{-2}$; this corresponds to our estimate based on *Miyake et al.* [2012]. Finally, *Usoskin and Kovaltsov* [2012] state that for a soft spectrum such as that associated with the 4 August 1972 SPE, the fluence should be about 1.2 x $10^{12}$ protons $cm^{-2}$ (40 times their estimated fluence given a 1956 SPE spectrum). We therefore take this as a third case. The October 1989 SPE is one of the most intense on record, and the measured E > 30 MeV proton fluence is 4.2 x $10^9$ [*McCracken et al.,* 2001]. Our three modelled cases correspond to 7.1, 71, and 285 times this 1989 fluence.

Figure 1 shows the globally averaged % change (comparing simulation runs with and without additional ionization input) in $O_3$ column density for the three SPE cases described above. We find maximum globally averaged depletions of 5%, 22% and 32% for the 1989-spectrum cases with fluence of 3 x $10^{10}$ and 3 x $10^{11}$ protons $cm^{-2}$, and the 1972-spectrum case with fluence 1.2 x $10^{12}$ protons $cm^{-2}$, respectively. In *Melott and Thomas* [2012] we considered the case of a short GRB and reported a maximum globally averaged $O_3$ depletion of about 10%, which is intermediate between the two lower fluence SPE cases considered here.  For comparison, current anthropogenic globally averaged depletion is about 3-5%.

In order to quantify surface-level impacts of UV under the ozone-depleted atmosphere we have used version 4.6 of the publically available Tropospheric Ultraviolet and Visible (TUV) atmospheric radiative transfer model, downloaded from http://cprm.acd.ucar.edu/Models/TUV/ [*Madronich and Flocke,* 1997].  Here we report computed values at 55° North latitude.  We find a maximum increase in UVB irradiance of about 30% for the low-fluence case, 160% for the mid-fluence case, and 317% for the high-fluence (soft spectrum) case.  These values correspond to maximum increases in erythema (skin damage due to solar UV) of 14%, 87%, and 160%, respectively.  Some research [e.g. *Björn et al.* 1996] has been done on UVB effects to land plants, and suggests possible disruption to plant growth and to interactions within ecosystems, but suggests that at least for current anthropogenic levels of depletion, agriculture should not be strongly impacted.  We have used a weighting function [*Flint and Caldwell* 2003] to compute UVB damage to terrestrial plants and find maximum increases at 55° North latitude of 2%, 14%, and 25%, respectively.

For the two lower fluence (hard spectrum) cases, these results suggest mild to moderate effects on the biota: some reduction of primary photosynthesis in the oceans and probably with land plants, increased risk of erythema and skin cancer [*Sweet et al.*, 2012], but no major mass-extinction level effects as implied by *Miyake et al.* [2012].  However, our results for a fluence of $1.2 \times 10^{12}$ protons cm$^{-2}$ with the soft August 1972 spectrum imply severe damage to the biosphere.  Given the lack of historical evidence for such damage, we can rule out such a high fluence that would be associated with a soft-spectrum event.

The main mechanism of recovery of the atmosphere following an ionization event is by incorporation of NO$_x$ into water (forming nitric acid) that is then rained or snowed out, usually deposited in ice cores. More details on our modelling of nitrate deposition can be found in *Melott et al.* [2005] and *Thomas et al.* [2007].  We have examined the GISP2 [*Mayewski et al.* 1997] ice core data, and find no large nitrate enhancements within 50 yr of their dating to 775. However, this data has a sampling interval of about 2.5 yr at the time in question, so that any period of nitrate enhancement may have been missed.  That question aside, this observation is consistent with our modelling, which predicts only about 2% enhancement in nitrate deposition at maximum, which would be down in the noise. Our conclusion using nitrate deposition combined with the $^{14}$C data (and previous GRB work [*Melott and Thomas*, 2012]) is that the event is consistent with either an SPE or a GRB in A.D. 774-775.

**5. Discussion**

There are at least three possible astrophysical sources for the observed 8$^{th}$ century $^{14}$C enhancement. First, a galactic GRB could be sufficiently distant, but still provide the necessary irradiation, so that its remnant might not have been observed. However, the *a priori* probability is quite small [*Melott and Thomas*, 2011, 2012; *Hambaryan and Neuhaeuser*, 2013]. We cannot evaluate the probability of a solar comet impact [*Eichler and Mordecai* 2013].

Data on the rates of CMEs with energy an order of magnitude or more greater than modern events such as October 1989 are lacking, and without spectral data to establish any firm pattern of fluences, but our energy estimates are consistent with the estimates and extrapolations of *Schrijver et al.* [2012] for events that occur every ~1000 years. It is consistent (and even possibly somewhat low) compared with recent observations of solar-type stars. Our simulations assuming a relatively hard spectrum and fluences consistent with estimates from *Miyake et al.* [2012] and *Usoskin and Kovaltsov* [2012] show moderate ozone depletion, which would have some deleterious effects on the biota for a few years, but is not a mass extinction level event. The soft spectrum case can be ruled out due to much more intense ozone depletion and subsequent biological impact. This conclusion regarding spectral hardness is in agreement with conclusions based on cosmogenic isotope production from *Miyake et al.* [2012], as well as *Hambaryan and Neuhaeuser* [2013] (who considered the case of a GRB). The enhancement to atmospheric nitrate deposition is small, consistent with its non-observation in the GISP2 ice core for the time period in question. The SPE appears to be the more reasonable possibility, which demands further exploration of a potential massive threat to modern civilization.

### 7. Acknowledgments

We thank Allen West for interesting discussions that led to this investigation, and two referees whose comments improved the paper. Computational time for this work was provided by the High Performance Computing Environment (HiPACE) at Washburn University; thanks to Steve Black for assistance with computing resources. Thanks to Sasha Madronich for assistance with using and modifying the TUV code.

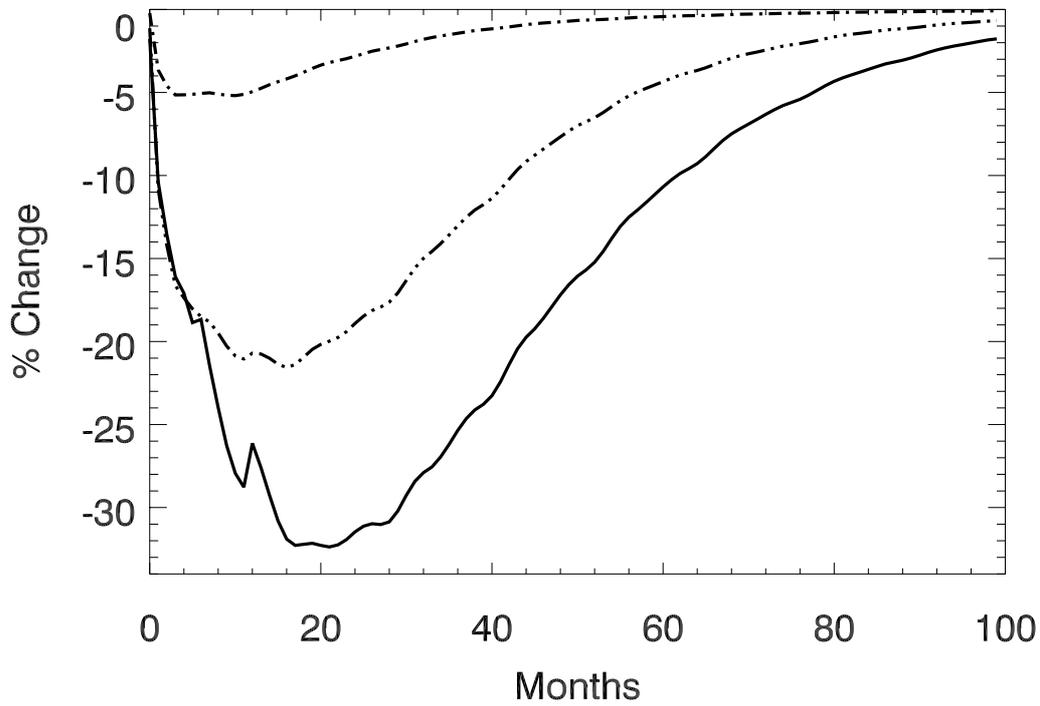

Figure 1 caption: Globally averaged percent change (comparing simulation runs with and without additional ionization input) in $O_3$ column density for 100 months after the event, for the three SPE cases considered here. Dot-dash line = October 1989 SPE spectrum with fluence $3 \times 10^{10}$ protons $cm^{-2}$; three-dot-dash line = October 1989 SPE spectrum with fluence $3 \times 10^{11}$ protons $cm^{-2}$; solid line = August 1972 SPE spectrum with fluence $1.2 \times 10^{12}$ protons $cm^{-2}$.